\documentclass[12pt,a4paper]{article}
\usepackage{float}
\usepackage{subfig}
\usepackage{amsmath}
\usepackage{amssymb}
\usepackage{graphicx}
\usepackage{hyperref}
\usepackage{csquotes}
\usepackage{multirow}
\usepackage[left=.45in,right=.45in,top=.6in,bottom=.6in,headheight=14.5pt]{geometry}
\usepackage{multirow}
\usepackage{pdflscape}
\usepackage[title]{appendix}
\usepackage[left=.45in,right=.45in,top=.6in,bottom=.6in,headheight=14.5pt]{geometry}
\usepackage{fmtcount} 
\usepackage{array,multirow}

\newcolumntype{C}[1]{>{\centering\arraybackslash}m{#1}}
\usepackage{epsfig}
\usepackage{float}
\usepackage{geometry}
\usepackage{amsmath}
\usepackage{cite}
\usepackage{ctable}

\newgeometry{vmargin={10mm,25mm},hmargin={19mm,19mm}}

\begin{document}

\title{\textcolor{blue}{Phenomenology of Vanishing Effective Majorana Mass with a Sterile Neutrino under Cosmological and JUNO Constraints}}

\author{Rushi Chambyal\thanks{rushichambyal@gmail.com }, Tapender \thanks{tapenderphy@gmail.com}, Labh Singh \thanks{sainilabh5@gmail.com}  and Surender Verma\thanks{s\_7verma@hpcu.ac.in}}

\date{%
Department of Physics and Astronomical Science\\
Central University of Himachal Pradesh\\
Dharamshala, India 176215
}
\maketitle
\begin{abstract}
\noindent In the present work we investigate the phenomenological implications of a vanishing effective Majorana neutrino mass  within a $3+1$ neutrino framework adding a eV-scale sterile neutrino beside three active neutrino states in light of latest cosmology driven bounds on sum of neutrino masses ($\sum_{i}m_i$). We explore the parameter space where the destructive interference between active and sterile states leads to vanishing amplitude, $M_{ee}$, of neutrinoless double beta ($0\nu\beta\beta$) decay. The allowed parameter space has been identified and predictions have been obtained taking into account the latest Planck and DESI+CMB bound on $\sum_{i}m_i$. We find that these bounds restrict the sterile mixing angle $\theta_{14}$ and the lightest active neutrino mass. Furthermore, we incorporate the refined precision data from JUNO experiment regarding solar oscillation parameters ($\theta_{12}, \Delta m_{21}^2$). We find that the sterile neutrino parameters like $\theta_{14}$ may not be sensitive to the JUNO precision measurements as the constraint imposed by precise $\theta_{12}$ is washed out by new cancellations driven through additional CP violating phases leading to vanishing $|M_{ee}|$.

\end{abstract}
\section{Introduction}

Over the past two decades solar, atmospheric, reactor and accelerator based neutrino oscillation experiments have established that neutrinos are massive and they mix non-trivially, thus, fixing the active neutrino parameters like mixing angles and mass-squared differences. These observations incontrovertibly established the evidence of new physics beyond the standard model (SM) of particle physics. The new experiments such as DUNE\cite{DUNE:2020ypp}, NO$\nu$A\cite{NOvA:2007rmc,NOvA:2023iam} and JUNO\cite{JUNO:2015zny,JUNO:2022mxj,JUNO:2025fpc}, to name a few, are assiduously striving to precise determine oscillation parameters. Yet several foundational questions still remain open such as the absolute neutrino mass scale, the ordering of the mass spectrum, the origin of neutrino mass, the nature of neutrinos (Dirac or Majorana), and the possible existence of sterile neutrinos.  Each of these issues points to an extension of the SM and motivates more detailed theoretical and experimental investigations. In particular, there exist intriguing anomalies posed by Liquid Scintillator Neutrino Detector (LSND)\cite{LSND:1995lje,LSND:2001aii}, MiniBooNE\cite{MiniBooNE:2007uho,MiniBooNE:2010idf}, reactor\cite{Huber:2011wv} and gallium\cite{Kaether:2010ag} anomalies which can be resolved by introducing new eV scale massive neutrinos called ``sterile neutrino", having no SM interactions \textit{i.e.} are singlet under the SM gauge group. There mixing with the active neutrino flavors are strongly constrained but not disallowed\cite{Gariazzo:2015rra}. This provides a unique opportunity as inquire about their existence as they can now manifests in the neutrino oscillation and direct neutrino mass measurement experiments. Moreover, they will have imperative implication in the dark matter searches and in cosmological scenarios affecting the large scale structure formation. On the theoretical front, sterile neutrinos naturally appear in frameworks explaining small neutrino mass including seesaw extensions, gauge-extended models, and flavor-symmetric constructions.

\noindent Apart from short baseline anomalies discussed above, if one allows for the dimension-5 lepton number violation and considering neutrinos to be Majorana particle allows for unique signature process \textit{aka} neutrinoless double beta ($0\nu\beta\beta$) decay.  Its observation would confirm lepton number violation and establish the Majorana nature of neutrinos. The decay rate depends on the effective Majorana mass $|M_{ee}|$, which encapsulates contributions from all neutrino mass eigenstates weighted by mixing angles and Majorana phases. The experiments  nEXO\cite{nEXO:2021ujk}, LEGEND\cite{LEGEND:2017cdu}, CUORE/CUPID\cite{CUORE:2011boi,CUORE:2019yfd}, KamLAND-Zen\cite{KamLAND-Zen:2012mmx}, and MAJORANA DEMONSTRATOR\cite{Majorana:2019nbd} are some experiments which are searching for this process but no conclusive signal has been observed so far. The possibility of existence of sterile neutrino may, further, obscure this picture because new mass eigenstate and CP violation phases will provide novel cancellations which may result in vanishing $|M_{ee}|$. The study of vanishing elements in the neutrino mass matrix, $i.e.$, texture zeros, has been extensively explored in the literature \cite{Xing:2002ta, Xing:2002ap, Kageyama:2002zw, Dev:2006qe, Ludl:2011vv, Kumar:2011vf, Fritzsch:2011qv, Meloni:2012sx, Meloni:2014yea, Dev:2014dla, Dev:2015lya, Borah:2015vra, Kaneko:2002yp, Kaneko:2003cy, Dev:2010vy, Bando:2004hi, Nguyen:2014mwa, Kalita:2015tda, Bora:2016ygl, Kitabayashi:2018bye,Borgohain:2019pya,Singh:2022ijf,Singh:2022tvz,Raj:2024dph}.

 \noindent The anisotropies in the cosmic microwave background radiation (CMBR), as measured by the Planck satellite, place a stringent upper bound on the sum of neutrino masses, $\sum_i m_i < 0.12~\text{eV}$ at 95\% confidence level (CL)~\cite{Planck:2018vyg}. More recently, an even tighter constraint, $\sum_i m_i < 0.072~\text{eV}$ at 95\% CL, has been obtained by combining Planck CMB data with baryon acoustic oscillation (BAO) measurements from the Dark Energy Spectroscopic Instrument (DESI)~\cite{DESI:2024mwx}. The sterile states may leave imprints on the expansion rate and structure formation. The recent cosmological analyses as stated above particularly the strong upper bound $\sum m_i < 0.072$ eV tightly restrict the allowed mass scales as demanded by oscillation-driven sterile neutrino hints. 
 This conflation places sterile-neutrino models to test as they remain compelling from a theoretical standpoint, yet their parameter space may increasingly be constrained by cosmological observations.

\noindent Furthermore,  current neutrino oscillation experiments are entering a precision era. JUNO’s high-resolution measurements of solar mixing angle $\theta_{12}$ and mass-squared difference, $\Delta m_{21}^2$ may further improve constraining the allowed parameter space of active–sterile mixing.  Although the possibility of vanishing $0\nu\beta\beta$ decay amplitude $|M_{ee}|$ has been extensively studied \cite{Verma:2016jpb,Li:2011ss,Giunti:2015kza,Borah:2015ufa,Dekens:2020ttz,Deepthi:2019ljo,Vien:2021hha,Ankush:2021opd,Fang:2024hzy,Priya:2025avk,Jana:2024xmc,Goswami:2026qpl} but the recent cosmology driven neutrino mass bounds and JUNO's high precision measurements demands a revisit of $(3+1)$ framework achieving vanishing effective Majorana mass $|M_{ee}|$. In the present work, we, by incorporating the most stringent cosmological bound and the latest JUNO-motivated oscillation inputs, we explore a tightly restricted yet phenomenologically rich region of the parameter space involving the lightest neutrino mass, the sterile mixing angle $\theta_{14}$, and the Majorana phases. The Monte-Carlo sampling techniques allow us to map the correlations among these parameters and identify viable solutions consistent with the cancellation in $|M_{ee}|$.

\noindent In Sec.~\ref{sec2}, we present the formalism and parameterization of the $(3+1)$ mass matrix. Sec.~\ref{sec3} contains the numerical methodology and a detailed discussion of the correlation plots for both normal and inverted mass hierarchies. Sec.~\ref{sec4} summarizes the main findings, implications for future experiments and the overall viability of the sterile neutrino in $3+1$ framework in light of DESI+CMB and JUNO data.

\section{Formalism}\label{sec2}


In the presence of one sterile neutrino, the leptonic mixing matrix 
is a $4\times 4$ unitary matrix $U$, which relates the flavor fields 
$\nu_\alpha = (\nu_e,\nu_\mu,\nu_\tau,\nu_s)^T$ 
to the mass eigenstate fields 
$\nu_i = (\nu_1,\nu_2,\nu_3,\nu_4)^T$ \textit{via},
\begin{equation}
\nu_\alpha = U_{\alpha i} \, \nu_i .
\end{equation}
For Majorana neutrinos, the neutrino mass matrix in the flavor basis is given by
\begin{equation}
M_\nu = U \, \mathrm{diag}(m_1, m_2, m_3 , m_4) \, U^T .
\end{equation}

\noindent We use the standard $(3+1)$ parametrisation \cite{Kang:2013zma} consisting of three 
active-active rotations and three active-sterile rotations along with 
Majorana phases. The mixing matrix may be written as
\begin{equation}
U = R_{34}(\theta_{34}) \,
    R_{24}(\theta_{24},\delta_{24}) \,
    R_{14}(\theta_{14},\delta_{14}) \,
    R_{23}(\theta_{23}) \,
    R_{13}(\theta_{13},\delta_{13}) \,
    R_{12}(\theta_{12}) \,
    P ,
\end{equation}
where the Majorana phase matrix is
\begin{equation}
P = \mathrm{diag}\left(1,\, e^{i\alpha},\, e^{i\beta},\, e^{i\gamma}\right),
\end{equation}
where $\alpha$, $\beta$, and $\gamma$ are Majorana CP phases and $\delta 's$ are Dirac CP phases. The rotation matrices are defined as
\begin{equation}
R_{12} = 
\begin{pmatrix}
c_{12} & s_{12} & 0 & 0 \\
-s_{12} & c_{12} & 0 & 0 \\
0 & 0 & 1 & 0 \\
0 & 0 & 0 & 1 
\end{pmatrix},
\qquad
R_{13} = 
\begin{pmatrix}
c_{13} & 0 & s_{13} e^{-i\delta_{13}} & 0 \\
0 & 1 & 0 & 0 \\
-s_{13} e^{i\delta_{13}} & 0 & c_{13} & 0 \\
0 & 0 & 0 & 1 
\end{pmatrix},
\end{equation}
\begin{equation}
R_{14} = 
\begin{pmatrix}
c_{14} & 0 & 0 & s_{14} e^{-i\delta_{14}} \\
0 & 1 & 0 & 0 \\
0 & 0 & 1 & 0 \\
-s_{14} e^{i\delta_{14}} & 0 & 0 & c_{14}
\end{pmatrix},
\end{equation}
and similarly for $R_{23}$, $R_{24}$, and $R_{34}$. Here,
\[
c_{ij}=\cos\theta_{ij},\qquad s_{ij}=\sin\theta_{ij}.
\]

\noindent The $(ee)$ element of $M_\nu$ is given by
\begin{equation}
(M_\nu)_{ee}
= m_1 c_{12}^2 c_{13}^2 c_{14}^2
+ m_2 s_{12}^2 c_{13}^2 c_{14}^2 e^{2i\alpha}
+ m_3 s_{13}^2 c_{14}^2 e^{2i\beta}
+ m_4 s_{14}^2 e^{2i\gamma}.
\end{equation}
Separating real and imaginary parts,
\begin{align}
\mathrm{Re}(M_{ee}) &= 
c_{12}^2 c_{13}^2 m_1
+ c_{13}^2 m_2 s_{12}^2 \cos(2\alpha)
+ m_3 s_{13}^2 \cos(2\beta)
- m_4 s_{14}^2 \cos(2\gamma),
\label{eq8}\\
\mathrm{Im}(M_{ee}) &= 
c_{13}^2 m_2 s_{12}^2 \sin(2\alpha)
+ m_3 s_{13}^2 \sin(2\beta)
- m_4 s_{14}^2 \sin(2\gamma).\label{eq9}
\end{align}

\noindent The sterile mixing angle $\theta_{14}$ enters only through $s_{14}^2$ in the 
expression for $M_{ee}$.  
We, therefore, solve Eqns. \eqref{eq8} and \eqref{eq9} for $s_{14}^2$. From the real part we get
\begin{equation}
s_{14}^2 = 
\frac{
c_{12}^2 c_{13}^2 m_1
+ c_{13}^2 m_2 s_{12}^2 \cos(2\alpha)
+ m_3 s_{13}^2 \cos(2\beta)
}{
c_{12}^2 c_{13}^2 m_1
+ c_{13}^2 m_2 s_{12}^2 \cos(2\alpha)
+ m_3 s_{13}^2 \cos(2\beta)
- m_4 \cos(2\gamma)
}
\equiv s_{14,sa}.
\end{equation}

\noindent Similarly, from the imaginary part
\begin{equation}
s_{14}^2 =
\frac{
c_{13}^2 m_2 s_{12}^2 \sin(2\alpha)
+ m_3 s_{13}^2 \sin(2\beta)
}{
c_{13}^2 m_2 s_{12}^2 \sin(2\alpha)
+ m_3 s_{13}^2 \sin(2\beta)
- m_4 \sin(2\gamma)
}
\equiv s_{14,sb}.
\end{equation}

\noindent A physical solution must satisfy
\begin{equation}
s_{14,sa} \simeq s_{14,sb},
\end{equation}
within a tolerance defined as $\mathcal{O}(10^{-4})$. In fact, in the numerical scan we impose the relative accuracy condition
\begin{equation}
\left|
\frac{s_{14,sa} - s_{14,sb}}{s_{14,sa}}
\right| < 10^{-4}.
\end{equation}
If this condition is satisfied, a physical value is assigned as $s_{14} = \sqrt{s_{14,sa}}$.
\noindent For the normal hierarchy (NH), the masses are sampled as
\begin{align}
m_1 &\in [0,\,0.05]\ \mathrm{eV}, \\
m_2 &= \sqrt{m_1^2 + \Delta m_{21}^2} ~\mathrm{eV}, \\
m_3 &= \sqrt{m_2^2 + \Delta m_{31}^2} ~\mathrm{eV}, \\
m_4 &= \sqrt{m_1^2 + \Delta m_{41}^2} ~\mathrm{eV},
\end{align}

\noindent while for inverted hierarchy (IH), the masses are sampled as
\begin{align}
m_3 &\in [0,\,0.05]\ \mathrm{eV}, \\
m_1 &= \sqrt{m_3^{\,2} + \Delta m_{23}^{2} - \Delta m_{12}^{2}}~ \mathrm{eV},\\
\qquad
m_2 &= \sqrt{m_1^{\,2} + \Delta m_{12}^{2}}  ~\mathrm{eV},\\
\qquad
m_4 &= \sqrt{m_1^{\,2} + \Delta m_{41}^{2}} ~\mathrm{eV}.
\end{align}
The scanned parameter set is thus
\begin{equation}
\{ m_1 (m_3),\alpha,\beta,\gamma,
\Delta m_{21}^2,\Delta m_{31}^2,
\Delta m_{41}^2, s_{12}^2, s_{13}^2 \}.
\end{equation}
In the numerical analysis we have followed the following steps: (i) Random uniform sampling of the first four parameters \textit{viz.,} the lightest mass and Majorana phases (ii) Gaussian sampling of the rest five oscillation parameters (iii) Consistency testing to obtain $s_{14}$ to obtain physically viable parameter sets. We, also, calculate sum of neutrino masses $\sum_i m_i = m_1 + m_2 + m_3$, which is relevant for cosmological constraints. 

\begin{table}[t]
\small
    \centering
    \renewcommand{\arraystretch}{1} 
    \begin{tabular}{ccccc} 
        \hline
        Parameter & best-fit$\pm1\sigma$ range & best-fit$\pm1\sigma$ range & $3\sigma$ range (NH) & $3\sigma$ range (IH) \\
 & (NH)& (IH)& &\\
        \hline
        $\sin^2\theta_{12}$ & $0.308^{+0.012}_{-0.011}$ & $0.308^{+0.012}_{-0.011}$ & $0.275 - 0.345$ & $0.275 - 0.345$ \\
        $\sin^2\theta_{23}$ & $0.470^{+0.017}_{-0.013}$ & $0.562^{+0.012}_{-0.015}$ & $0.435 - 0.585$ & $0.410 - 0.623$ \\
        $\sin^2\theta_{13}$ & $0.02215^{+0.00056}_{-0.00058}$ & $0.02224^{+0.00056}_{-0.00057}$ & $0.02023 - 0.02388$ & $0.02053 - 0.02397$ \\
        $\frac{\Delta m^2_{3l} } {10^{-3} \text{eV}^2}$& $2.513^{+0.021}_{-0.019}$ & $-2.510^{+0.024}_{-0.025}$ & $2.463 - 2.606$ & $-2.584 - -2.438$ \\
        $\frac{\Delta m^2_{21}} {10^{-5} \text{eV}^2}$& $7.49^{+0.19}_{-0.19}$ & $7.49^{+0.19}_{-0.19}$ & $6.92 - 8.05$ & $6.92 - 8.05$ \\
        \hline
    \end{tabular}
    \caption{The values of the active neutrino parameters as given by NuFIT 6.0~\cite{Esteban:2024eli} used in the numerical analysis.}
    \label{tab:nudata}
\end{table}

\begin{table}[t]
\centering
\small
    \centering
    \renewcommand{\arraystretch}{1} 
  
\centering
\begin{tabular}{l c}
\hline
Parameter & Range used in this work  \\
\hline
$\sin^2\theta_{14}$ & $0.0098 - 0.031$ \\

$\Delta m^2_{41}\,(\mathrm{eV}^2)$ & $0.35 - 2$ \\
\hline
\end{tabular}
\caption{The sterile neutrino oscillation parameters used in this work for both normal hierarchy (NH) and inverted hierarchy (IH), consistent with current experimental bounds\cite{Gariazzo:2017fdh}.}
\label{tab:sterile_parameters}
\end{table}

\section{Numerical Analysis and Discussion}\label{sec3}

\noindent In the correlation plots shown in Fig.~1(a–f), we present numerical predictions of a $3+1$ sterile–neutrino mass–matrix framework with vanishing $|M_{ee}|$ under the normal hierarchy focusing on the interplay among the lightest neutrino mass $m_{1}$, the active–sterile mixing parameter $\sin\theta_{14}$, the Majorana phases $(\alpha,\beta,\gamma)$, and the cosmological constraint $\sum m_i$. The correlation plots shown in Fig. 1 are at 3$\sigma$ CL. The region of the parameter space represented by  blue  points refers to vanishing effective Majorana mass $|M_{ee}|$ consistent with NuFit 6.0 data shown in Table 1. The orange points represents the allowed parameter space considering, in addition, the cosmological bound of 0.12 eV on $|\sum_i m_i|$ \cite{Planck:2018vyg} and global-fit range of $\sin\theta_{14}$\cite{Gariazzo:2017fdh}. It is evident from Figs. 1(a-d) that the Majorana phases $\alpha, \beta$ and $\gamma$ remains unconstrained except the point $\gamma\approx90^o$.  
While the predictions for the Majorana phases $\alpha$ and $\beta$ remain unchanged, imposing the cosmological bound together with the global fit range of $\sin \theta_{14}$ significantly restricts the allowed range of $\gamma$ to $(70^\circ$–$110^\circ)$. The global fit range of the sterile mixing angle $\sin\theta_{14}$ shown by vertical dashed lines in Fig. 1(c) has interesting implication for Majorana phase $\alpha$. It is evident that near the lower values of $\sin\theta_{14}$ in the range, whole range of $\alpha$ is allowed, however, as one go from lower to higher values of $\sin\theta_{14}$ in the range, there exist a region about $\alpha \approx 90^o$ which may become disallowed. For example, near $\sin\theta_{14}=0.15$, the range $80^o\le\alpha\le 100^o$ is disallowed. This conclusion remains intact even after the incorporation of cosmological bound on sum of neutrino masses in the analysis except that disallowed range now widens. The scenario of vanishing effective Majorana mass predicts $\sin\theta_{14}$ in the range ($0.01-0.23$) and an unconstrained $\gamma$ for $\sin\theta_{14}$ less than 0.07 as seen from Fig. 1(d). But Planck upper bound on $\sum_i m_i$ (0.12 eV) and current 3$\sigma$ range of $\sin\theta_{14}$ restricts $\gamma$ to ($70^o-110^o$). Figs. 1(e) and 1(f) demonstrate the correlation of $\sum_i m_i$ and $m_1$ with $\sin\theta_{14}$ respectively. It is evident from these figures that large part of the parameter space become disallowed for the conjecture of vanishing effective Majorana mass to be consistent with cosmological upper bound of 0.12 eV and current 3$\sigma$ range of $\sin\theta_{14}$. Moreover, an important outcome of the present analysis is that, upon adopting the more restrictive DESI+CMB upper bound of 0.072 eV, the parameter $\sin\theta_{14}$ must lie in the range ($0.10-0.13$) thus predicting an upper bound on $\sin\theta_{14}$. If future cosmological data tighten $\sum_i m_i$ below the DESI+CMB line, the region with $\sin\theta_{14} \gtrsim 0.1$ may be entirely excluded, forcing either extremely small sterile mixing or new physics that modifies early-universe neutrino evolution. 
\noindent 
 \begin{figure}[!h]
     \centering
     \includegraphics[width=0.7\linewidth]{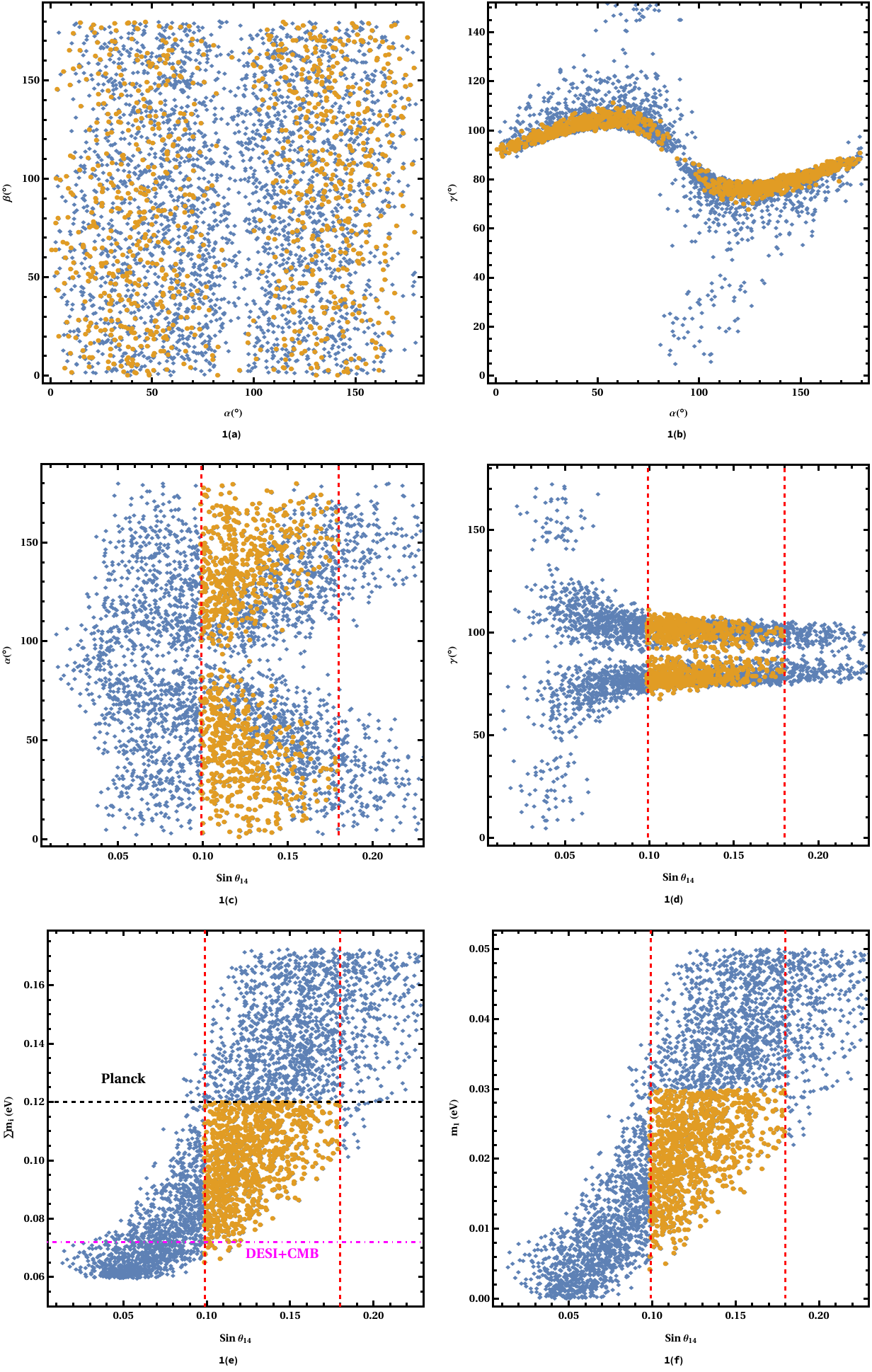}
     \caption{The correlation plots for NH of neutrino masses. The vertical dashed (red color) in Figs.~1(c-f) are the global-fit 3$\sigma$ range of $\sin\theta_{14}$. The horizontal dashed line (black color) in Fig.~1(e) is the Planck's upper bound on $\sum_i m_i$ while dot-dashed (magenta color) is the upper bound from DESI+CMB.}
     \label{fig:placeholder}
 \end{figure}

 \begin{figure}
     \centering
     \includegraphics[width=0.7\linewidth]{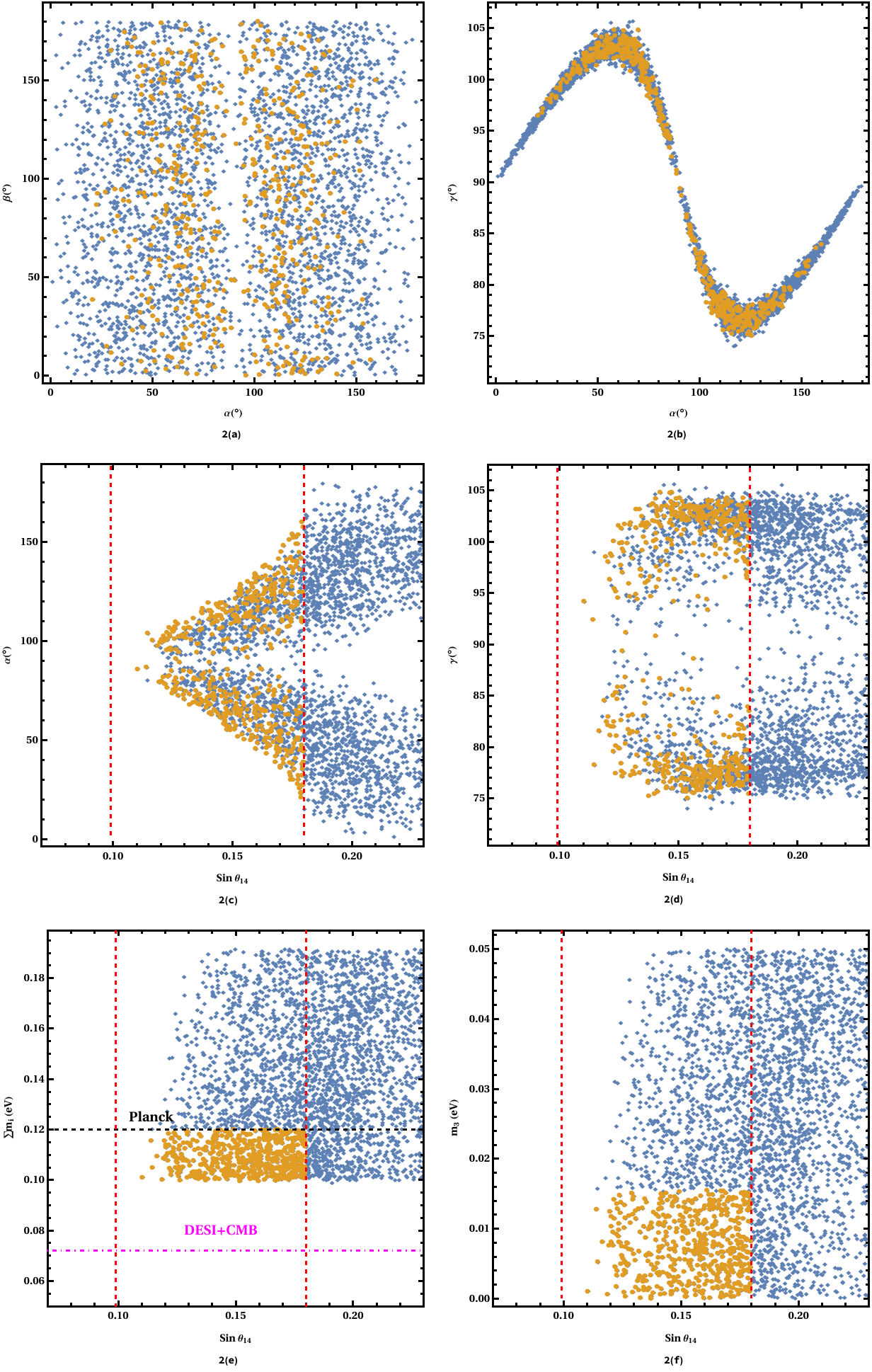}
     \caption{ The Correlation plots for IH of neutrino mass. The vertical dashed (red color) in Figs.~2(c-f) are the global-fit 3$\sigma$ range of $\sin\theta_{14}$. The horizontal dashed line (black color) in Fig.~2(e) is the Planck's upper bound on $\sum_i m_i$ while dot-dashed (magenta color) is the upper bound from DESI+CMB.}
     \label{fig:placeholder}
 \end{figure}

\noindent Figs.~2(a–f) exhibit the correlation plots amongst the same parameters for inverted hierarchical neutrino masses. The correlation plot between Majorana phases $\alpha$ and $\beta$ are similar to that of NH case, for example, full ranges of the phases $\alpha$ and $\beta$ are allowed except the point $\alpha \approx 90^o$. The  Majorana phase $\gamma$ is constrained to a relatively smaller range of ($75^o-105^o$) in IH case and $\gamma \approx 90^o$ is disallowed. One of the important consequence of vanishing $|M_{ee}|$ in $3+1$ framework is the existence of lower bound on $\sin\theta_{14}$ as can be seen in Figs.~2(c-d). It is to be noted that the global-fit range of $\sin\theta_{14}$ is $(0.098 - 0.018)$ at 3$\sigma$ CL \cite{Gariazzo:2017fdh}, however, the model predicts a lower bound of $\sin\theta_{14}>0.115$, thus, pruning a considerable region of allowed parameter space. It is a generic bound and is not a consequence of the cosmological upper bound on sum of neutrino masses. Furthermore, the model predicts a lower bound on $\sum_i m_i>0.1\;\mathrm{ eV}$. So, the model has high  predictive power in the IH with $\sum_i m_i$ constrained to a range ($0.10-0.12\;\mathrm{eV}$), where upper bound is coming from cosmological analysis of the formation of large scale structures and lower bound is driven by the model itself. Importantly, if one considers the more severe recent bound on $\sum_i m_i$ from DESI+CMB, IH is ruled out in the current scenario, however, as we discussed previously NH is allowed as a small region is still allowed for $\sin\theta_{14}<0.13$ (Fig.~1(e)).

\noindent From a phenomenological point of view, these results carry significant implications \textit{viz.,} NH still accommodates a broad region of sterile-mixing parameter space, IH is already under notable pressure, and forthcoming next generation cosmological surveys (for example, CMB-S4 and DESI extensions) could effectively eliminate the remaining allowed region if they push the $\sum_i m_i$ bound even slightly downward. At the same time, future oscillation experiments such as JUNO \cite{JUNO:2025fpc}, DUNE \cite{DUNE:2020ypp}, and Hyper-Kamiokande \cite{Hyper-Kamiokande:2018ofw,Hyper-Kamiokande:2025fci} will sharpen sensitivity to active–sterile mixing and may be capable of resolving $\sin\theta_{14}$ at the percent level, which may have imperative implications for both NH and IH and could decisively test the viability of vanishing $|M_{ee}|$ in $3+1$ scenario. 
\begin{figure}
\centering
\begin{tabular}{cc}
     \includegraphics[width=0.5\linewidth]{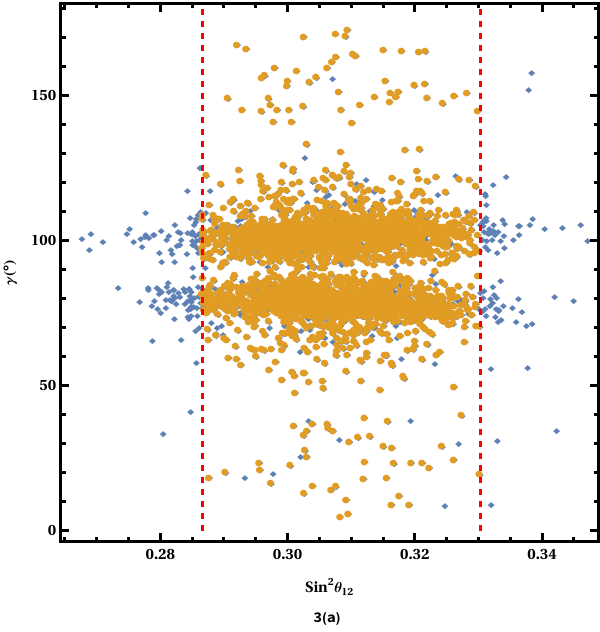} &
     \includegraphics[width=0.5\linewidth]{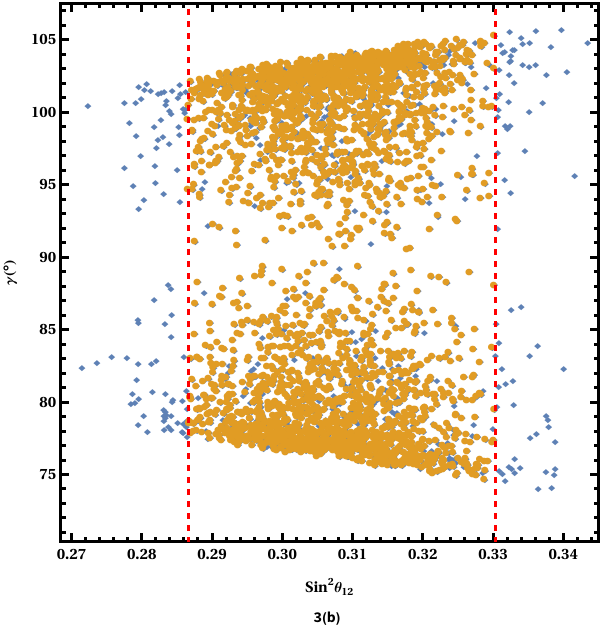}
\end{tabular}
     \caption{Correlation between $\sin^2\theta_{12}$ and the phase $\gamma$ for normal hierarchy (NH) and inverted hierarchy (IH). The vertical dashed lines represent the SNO+ and JUNO $3\sigma$ constraint on $\sin^2\theta_{12}$.}
     \label{fig:placeholder}
 \end{figure}
 
\noindent We, also, explored the implications of the latest  measurement of  solar mixing angle $\theta_{12}$ by JUNO. The representative correlations are shown in Fig. (3). Due to lack of a correlation of parameters of the model with $\theta_{12}$ JUNO's high precision measurement may not significantly affect the parameter space except of Majorana phase $\gamma$, particularly, for inverted hierarchy. It is evident from Fig. 3(b), that the allowed range of the phase $\gamma$ increases  as one go from the lower to the upper values of gamma (excluding $\gamma=0^o$) within 3$\sigma$ range, however, no such correlation exists in Fig. 3(a) for normal hierarchy.

\section{Conclusions}\label{sec4}
\noindent In summary, we have revisited $3+1$ neutrino framework wherein one eV-scale sterile neutrino has been added to the fermionic content of the SM. We determine the allowed parameter space consistent with the condition of vanishing effective Majorana mass $|M_{ee}|$ in light of Planck, DESI+CMB and the latest JUNO data. The updated analysis clearly signify that cosmological bound on $\sum_i m_i$ inspired from Planck results exclude significant region of the parameter space (allowed parameter space has been shown in orange color in Figs.~(1-2)). Moreover, if we employ the more severe DESI+CMB bound then IH is ruled out in the model, thus, $|M_{ee}|$ can only vanish in NH case. In the NH, DESI+CMB bound, also, set an upper bound active-sterile mixing parameter $\sin\theta_{14}<0.13$. This cosmology driven bound is the most important prediction of the present analysis. Recently, KATRIN experiment has set value of the active-sterile mixing at percent level\cite{KATRIN:2025lph}. This again will rule out IH in the model, however, NH may still be allowed provided future cosmological upper bound of $\sum_i m_i$ becomes more stringent and shifts to the region around $\approx$ 0.06 eV.  The generic,  prediction of the model is the existence of lower bound on $\sin\theta_{14}>0.115$ which is consistent with the earlier analysis in Ref.\cite{Verma:2016jpb}. It is to be noted that the distinctive sensitivity makes IH simultaneously more predictive and more fragile within the $3+1$ scheme. Future CMB and large scale structure experiments along with precision oscillation facilities will decisively test the remaining allowed regions. 
Overall, the model remains viable, but only within tightly confined regions highlighting the critical role of future experimental sensitivities in establishing or ruling out sterile-neutrino extensions in this scenario.

\end{document}